\documentclass[twocolumn, superscriptaddress, aps, pra]{revtex4-2}
\usepackage[english]{babel}
\usepackage[utf8]{inputenc}
\usepackage[colorinlistoftodos, color=green!40, prependcaption]{todonotes}
\usepackage{multirow}
\usepackage[normalem]{ulem}
\usepackage{amsmath}
\usepackage{url}

\usepackage{siunitx}
\usepackage[pdftex, pdftitle={Article}, pdfauthor={Author}]{hyperref}

\newcommand\kay{\mathcal{k}}

\newcommand{\stateZeroDefAlt}{$|\Psi_1\rangle = \frac{1}{\sqrt{6}} \sum_{\ell=0}^5 |\ell\rangle$}
\newcommand{\stateThreeDef}{$|\Psi_2\rangle = \frac{1}{2} (|0\rangle + i | 2\rangle -i | 3\rangle  -| 4\rangle)$}

\usepackage{graphicx}
\usepackage{dcolumn}
\usepackage{bm}

\begin{document}

\title{Parallelized projective measurements for spatial photonic qudits estimation} 
\author{Quimey Pears Stefano}
\email{quimeymartin.pearsstefano@ehu.eus}
\affiliation{Centro de Física de Materiales (CFM), CSIC-UPV/EHU, Paseo Manuel de Lardizabal 5, 20018 Donostia-San Sebastián, Spain}

\author{Lorena Reb\'on}%
\affiliation{Instituto de F\'isica de La Plata, CONICET-UNLP, Diagonal 113 entre 63 y 64, La Plata (1900) - Buenos Aires - Argentina \\}
\affiliation{Departamento de Ciencias B\'asicas, Facultad de Ingenier\'ia UNLP, 1 y 47 La Plata, Argentina.\\}

\author{Claudio Iemmi}%
\affiliation{Universidad de Buenos Aires, Facultad de Ciencias Exactas y Naturales, Departamento de Física, Pabellón I, Ciudad Universitaria (1428), Buenos Aires, Argentina \\}
\affiliation{CONICET - Universidad de Buenos Aires, Buenos Aires, Argentina.\\}

\date{\today}

\begin{abstract}
We present a quantum state tomography method that enables the reconstruction of
\emph{arbitrary} $d-$dimensional quantum states encoded in the discretized transverse momentum of photons, by using \emph{only} $d+1$ experimental settings. To this end, we identify a family of bases with the property that the outcomes of a projective measurement are \emph{spatially multiplexed} on the interference pattern of the projected state. Using the proposed scheme we performed, as a proof-of-principle, an experimental reconstruction of $d=6-$dimensional states, for which a complete set of mutually unbiased bases does not exist. We obtained fidelity values above 0.97 for both pure and mixed states, reducing the number of experimental settings from $42$ to only $7$.
\end{abstract}


\maketitle

The determination of an unknown quantum state is a problem of both fundamental and applied interest in quantum information processing \cite{Paris2004}. Since typical quantum-state tomography methods for $d$-dimensional quantum states (\emph{qudits}) require a number of measurement settings that increases as $d^2$ \cite{Wootters1989,Adamson2010,James2001}, the search for schemes that reduce the number of measurements is of paramount importance for experiments involving high-dimensional states \cite{Zahidy2024, Farkas2021, Akatev2022}

Optimizing quantum measurement strategies is a key approach to improve the efficiency of quantum state tomography (QST). In particular, it can be significantly enhanced by a smart choice of bases, thereby reducing the number of measurements required to reconstruct the desired unknown state.
For example, the set of measurement bases proposed in Ref.~\cite{PearsStefano2019}, defines a QST scheme for \textit{pure} qudits that requires $4d$ measurement outcomes and can be parallelized, especially for spatially-encoded photonic qudits. This possibility of parallel measurements reduces the number of specific configurations of the experimental setup that are needed, to a total of only $4$, regardless of the dimension $d$ of the quantum system~\cite{PearsStefano2017}. On the other hand, for an arbitrary qudit (whose density matrix is not pure), typical schemes require of the order of $d^2$ projective measurements \cite{Thew2002}. In the case of spatially-encoded photonic qudits, each of these $d^2$ projections requires an individual measurement configuration, where the photocounts are measured at a fixed position in the far field \cite{Lima2011}.

In this article, we propose a quantum-state estimation scheme that allows us to reconstruct an arbitrary photonic qudit state, which is encoded in the discretized transverse momentum of the photon, using only $d+1$ experimental measurement configurations. The core of the method is a recipe to generate a family of bases, each of which has the property that the probability projections of a given arbitrary state \emph{onto all} its elements, can be measured simultaneously at $d$ positions on the far field. Thus, the result of a quantum projection onto any of these bases is spatially multiplexed in a similar way to what happens for polarization-encoded states through a polarizing multi-port beam splitter.
From this family of basis an informationally complete set can be generated, thus allowing to perform a full tomography of an arbitray qudit using only $d+1$ experimental configurations.

We will start by briefly reviewing the formalism of \emph{slit states}. Let us consider an array of $d$
slits of width $a$, where the separation between adjacent
slits is $s$. Since the length of these slits is $L (\gg a,s)$ and each of them has an arbitrary complex transmission $c_k$, we can described such an aperture by $A(\mathbf{x}) = \mathrm{Rect}(x/L) \sum_{\ell=0}^{d-1} c_\ell \;\mathrm{Rect}\left(\frac{y-\ell s}{a}\right)$,
being $\mathbf{x} =(x,y)$ the transverse position coordinate, and $\mathrm{Rect}(\eta)$ the 
rectangle function which takes the value 1 if $|\eta|< 1/2$, or 0 otherwise.
Hence, if a paraxial and monochromatic single-photon field, described by the normalized transverse probability amplitude $\psi(\mathbf{x})$, goes through this aperture, the resulting quantum state is 
\begin{align}\label{eq:spatial-qudit}
  |\Psi\rangle  = \int  
  A(\mathbf{x}) \psi(\mathbf{x})| 1\mathbf{x}\rangle \; \mathrm{d}\mathbf{x},
\end{align}
where
$| 1\mathbf{x}\rangle$ is the single-photon state in the position
basis. Therefore, if $\psi(\mathbf{x})$ is assumed to be approximately constant across the region of the slits, the quantum state described
by \eqref{eq:spatial-qudit} is the qudit state $|\Psi\rangle = \sum_{\ell=0}^{d-1} c_\ell | \ell\rangle$, 
where $\left\{| \ell\rangle\right\}_{\ell=0}^{d-1}$ is the logical basis. Then, how a projective measurement can be physically implemented on this kind of spatial photonic qudits is detailed in Ref.~\cite{Lima2011}. In that work, Lima et. al. show that the information of the projection of the state $|\Psi\rangle$ onto another qudit $|\Phi\rangle = \sum_{\ell=0}^{d-1} b_\ell | \ell\rangle$, is encoded in the far field interference pattern of $d$ slits with complex transmissivities $\left\{c_\ell b_\ell^* \right\}_{\ell=0}^{d-1}$.
In fact, since the interference pattern in the focal plane of a convergent lens of focal distance $f$ is given by 
\begin{align}\label{eq:interference-pattern}
 I(x) = \mathrm{sinc}^2\left(\frac{k x a}{2f}\right) \left|\sum_{\ell=0}^{d-1} c_\ell b_\ell^* \exp\left(i \ell \frac{ s k x}{f}\right)\right|^2,
\end{align}
with $k$ the wavenumber of the photon field, the intensity at the center will be proportional to the probability of projection between those two states: $I(x=0) \propto \left|\sum_{\ell=0}^{d-1} c_\ell b_\ell^* \right|^2 = |\langle \Phi|\Psi \rangle|^2.$

Equation~(\ref{eq:interference-pattern}) also allows us to infer that the intensity at \textit{any} position $x$ will be proportional to the probability of projecting $|\Psi\rangle$ onto a vector state $|\tilde{\Phi}\rangle = \sum_{\ell=0}^{d-1} \tilde{b}_\ell|\ell\rangle$, with $\tilde{b}_\ell = b_\ell \exp(i \ell s k x/f)$. 
We will show, in what follows, how a single pattern can be used to obtain significant information from the unknown state $|\Psi\rangle$, namely, to obtain its projection onto all the elements of a given basis. 

\textit{Multiplexing measurement along the interference pattern.}
Let us consider the interference pattern of an array of $d$ slits, at the particular positions given by 
\begin{align}\label{eq:interference-pattern-origin}
x^{(m)}  = f\frac{\lambda}{s} \frac{m}{d}, \qquad \mathrm{with}\,\,m=0,1,\dots, d-1,
\end{align}
where $\lambda = \frac{2\pi}{k}$ is the wavelength of the photon field. For $m\neq 0$, $x^{(m)}$ correspond to the minima of the interference
pattern of $d$ slits with equal amplitudes and phases, while $x^{(0)}$
corresponds to its central maximum \cite{Hecht}. At these positions, Eq.~(\ref{eq:interference-pattern}) can be rewritten as 
\begin{align}\label{eq:interference-pattern-minima}
 I(x^{(m)}) = \mathrm{sinc}^2\left(\pi\frac{a}{s}\frac{m}{d}\right) \left|\sum_{\ell=0}^{d-1} c_\ell b_\ell^* \omega^{m\cdot \ell}\right|^2,
\end{align}
where $\omega = \exp\left(i 2\pi /d\right)$ is the first $d$--root of unity. Hence, each of these $d$ positions along the interference pattern contains information about the projection of $|\Psi\rangle$ onto a different vector state $|\Phi_m\rangle$, being that%
\begin{align}\label{eq:basis}
 |\Phi_m\rangle = \sum_{\ell=0}^{d-1} b_\ell \omega^{m\cdot \ell} |\ell\rangle,
\,\,\,m=0,1,\dots, d-1.
\end{align}
\begin{figure}[!t]
\centering
\includegraphics[width=.45\textwidth]{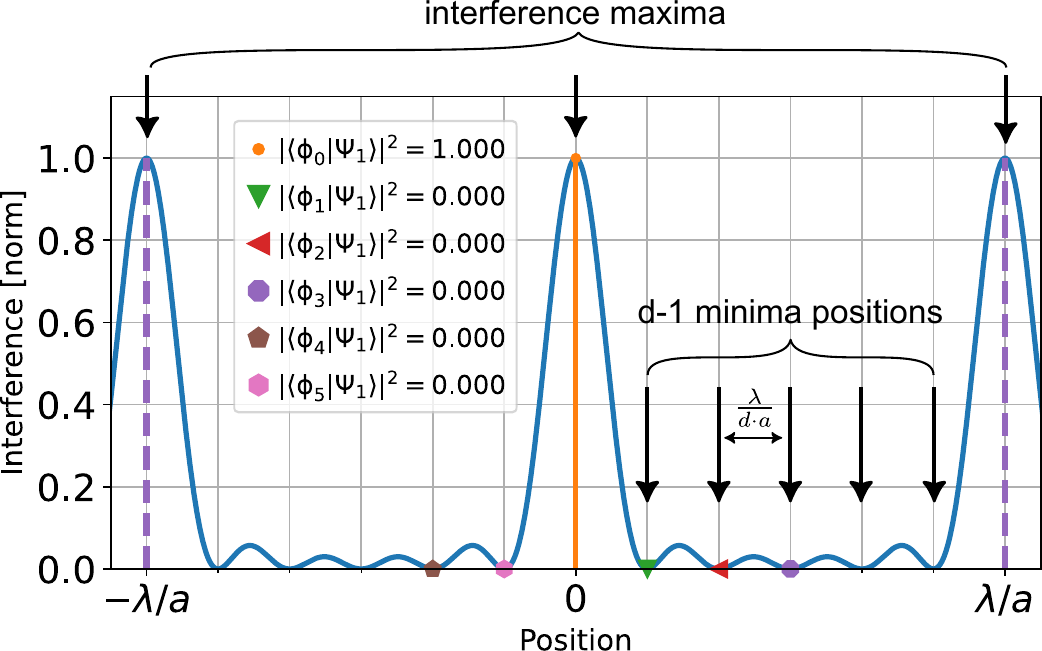}
\caption{Theoretical interference pattern corresponding to the projection of the state \stateZeroDefAlt\ onto the basis generated by itself, according to the rule $\{|\Phi_m\rangle\}_{m=0}^{5} = \{\frac{1}{\sqrt{6}}\sum_{\ell=0}^{5}  \omega^{m\cdot \ell} |\ell\rangle \}_{m=0}^{5}$. Each of the arrows indicates the position where the measured intensity gives the probability of $|\Psi_1\rangle$ over one of the elements of such a basis.} \label{fig:interference-patterns-theo}
\end{figure}
Furthermore, in the case where the condition $|b_\ell|^2 =1/d$ is fulfilled, for $\ell=0,\cdots,d-1$, i.e. all the slits in the array have equal transmissivities, it can be shown that such states form an orthonormal basis for qudits. Indeed, the interior product between any two of these states is%
\begin{align}\label{eq:multiplexed-bases}
 \langle \Phi_m|\Phi_n\rangle & = \sum_{\ell, \kay=0}^{d-1} b_\ell b_\kay^* \omega^{m\cdot \ell}\omega^{-n\cdot \kay} \langle\kay|\ell\rangle = \sum_{\ell,\kay=0}^{d-1} b_\ell b_\kay^* \omega^{m\cdot \ell -n\cdot \kay} \delta_{\kay,\ell} \nonumber\\
 & = \sum_{\ell=0}^{d-1} |b_\ell|^2 \omega^{(m-n)\cdot \ell} = \frac{1}{d} \sum_{\ell=0}^{d-1} \omega^{(m-n)\cdot \ell} = \delta_{m,n},
\end{align}
where we have used the fact that $\omega$ is a root of unity, and therefore the sum is not null only if $m=n$.

Thus, a set of $d$ states defined by \eqref{eq:basis}, is an orthonormal basis in a $d$--dimensional Hilbert space, and the projection probability of a qudit state onto \emph{all} the elements of such a basis can be estimated \emph{from a single interference pattern} of $d$ slits, whose complex transmissivities represent a given (but arbitrarily chosen) element of the basis.

As an example, but without loss of generality, we show in Fig.~\ref{fig:interference-patterns-theo} the theoretical interference pattern that represents the state \stateZeroDefAlt\  projected onto itself, i. e., the interference pattern of $d = 6$ slits with complex transmissivities, such that $c_i b_i^* = 1/6$ for $i=0,\dots,5$. From this simple case, it is clear to see that the (normalized) intensity value in the position of the maxima and minima, gives the projection probabilities onto the different elements of the basis $\{\frac{1}{\sqrt{6}}\sum_{\ell=0}^{5} \omega^{m\cdot \ell} |\ell\rangle\}_{m=0}^{5}$. Also, it is noteworthy that such probabilities can be recorded always in the same positions of the interference pattern, regardless of the state $|\Psi\rangle$ to be estimated.

\textit{Generating an informational complete set of bases.} To perform the tomographic reconstruction of an unknown quantum state, a complete set of $d+1$ bases, $\{\mathcal{B}_J\}_{J=0}^d$\;, can be generated as follows: (i) The first basis, $\mathcal{B}_0$, is chosen to be the canonical basis; 
(ii) for each of the next bases $(J=1,\cdots,d)$, a vector that fulfils the condition $|b_\ell|^2 =1/d$ $(\ell=0,\cdots,d-1)$ 
is selected at random, and the basis is completed according to \eqref{eq:basis};
(iii) the resulting linear system is solved after its condition number is evaluated to verify if it is a well-conditioned problem. Alternatively, this random generation of the set of bases can be repeated several times, then selecting the set with the best condition number.

Finally, having followed steps from (i) to (iii), we obtain a set of bases $\{\mathcal{B}_J\}_{J=0}^d$, where $\mathcal{B}_0 = \{|\ell\rangle\}_{\ell=0}^{d-1}$, and $\mathcal{B}_J = \{|\Phi_m^{(J)}\rangle\}_{m=0}^{d-1}$. This is a suitable set to univocally reconstruct the density matrix $\rho = \sum_{i,j} \rho_{i,j} |i\rangle\langle j|$ that describes an arbitrary quantum state (not only a pure one as given by a vector state like that of \eqref{eq:spatial-qudit}). Thus, the probability of projecting such state over the element $m$ of the $J-$basis is 
\begin{align}\label{eq:probability}
p_m^{(J)} = \mathrm{Tr} \left(\rho |\Phi_m^{(J)}\rangle\langle\Phi_m^{(J)}|\right) = \sum_{i,j} \rho_{i,j} (b_m^J)_j (b_m^J)^*_i,
\end{align} where $\mathrm{Tr}(\cdot)$ refers to the trace of a matrix, and $(b_m^J)_i$ is the $i$ component of the state $m$ of the basis $J$, $|\Phi_m^{(J)}\rangle$. The tomographic reconstruction can be performed if the matrix $M$, with elements $M_{i,j} = (b_m^J)_j (b_m^J)^*_i$, is non-singular~\cite{Thew2002}.

It should be noted that, except for the canonical basis $\mathcal{B}_0$, the projective measurements onto any of the bases in the generated informational complete set, can be spatially multiplexed on the positions specified by \eqref{eq:interference-pattern-origin}. Even more, as we will show in the experimental description, a customized diffraction grating can be used to spatially multiplexed the measurement in the canonical basis.
\begin{figure}[!ht]
\centering
\includegraphics[width=.45\textwidth]{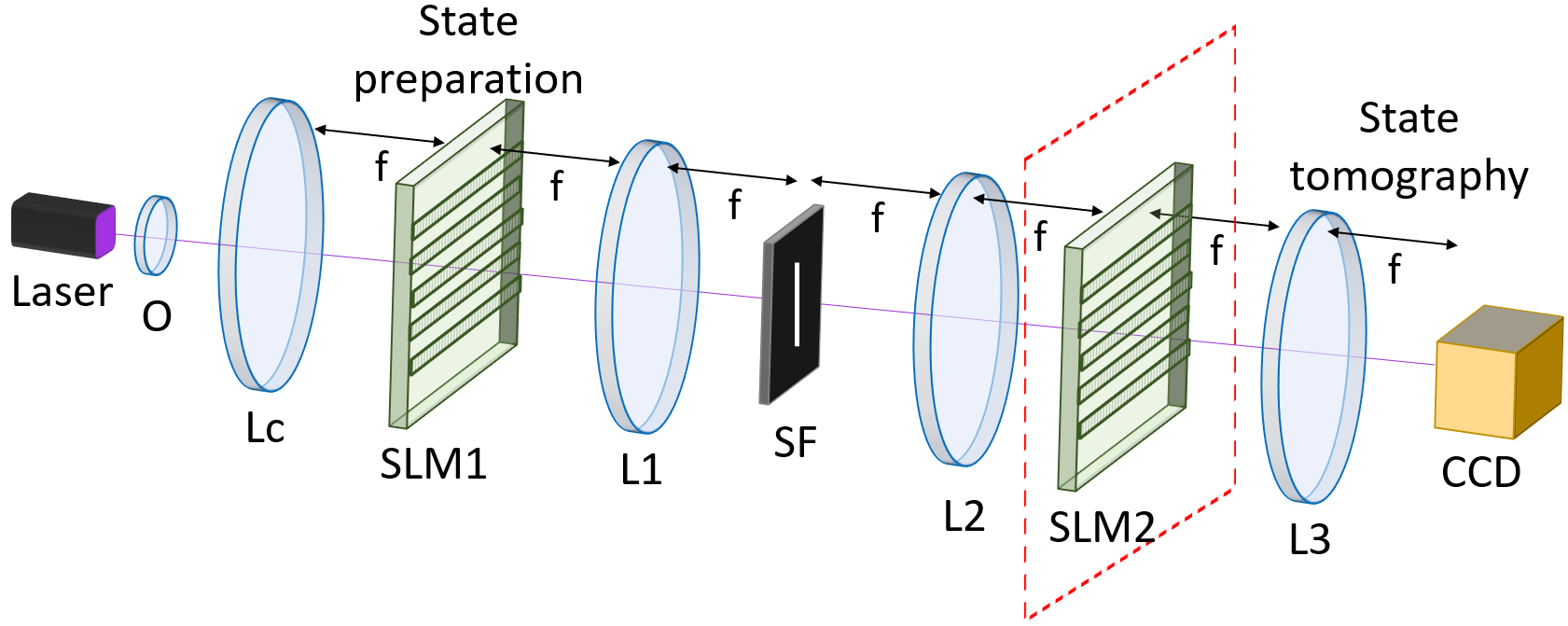}
\caption{Experimental setup. The light source is a 405 nm cw laser diode, expanded and collimated by the objective O and the convergent lens Lc, respectively. Lenses L1, L2 and L3 conform a convergent optical processor. SLM's are phase-only spatial light modulators and SF is a spatial filter. The full interferograms are detected by a CCD camera.\label{setup} } 
\end{figure}

\textit{Experimental implementation.} The setup used to implement the proposed tomographic method is sketched in Fig.~\ref{setup}. It can be divided into two modules: The first module is basically a $4$--$f$ optical processor which is used for the state preparation. A laser diode (405 nm), is expanded by a microscope objective (O), and then collimated by the lens (Lc). In this way a plane wave with almost constant intensity distribution illuminates the first spatial light modulator (SLM1), placed at the front focal plane of lens (L1). The SLM is conformed by a Sony liquid crystal television panel model LCX012BL which, in combination with polarizers and wave plates, provides the adequate state of polarization of light and allows a full phase modulation of the incident wavefront for the operating wavelength. An arbitrary slit state can be obtained at the back focal plane of lens (L2) following the method described in \cite{Solis-Prosser2013}. On the one hand, the absolute amplitude $|c_k|$ of each slit is controlled by means of a phase grating which diffraction efficiency depends on its phase modulation. To encode this information, we chose the first diffracted order, which is selected by the spatial filter (SF). On the other hand, the argument of $c_k$ is obtained by adding an adequate uniform phase on the grating modulation.
The second module performs the state tomography by means of SLM2, which provides the necessary phase modulation to encode the $m$--state of the basis $J$ ($|\Phi_m^J\rangle$), on which the projective measurement is carried out. Both the complex modulation encoding and the SLM specs, are the same as the used with the first module. SLM2 is placed at the front focal plane of lens (L3). Thus, the exact Fourier transform of the projected spatial qudit is obtained at the back focal plane of lens (L3). The light distribution corresponds to the interference pattern projection between the prepared state and the selected projector state described by \eqref{eq:interference-pattern}. The interference pattern was normalized such that the sum of the intensity values in the positions of interest is equal to 1, i. e., $p_m = I(x^{(m)})/\sum_j I(x^{(j)})$. Then, the obtained values correspond to the probability distribution of the projection of the target state over the given basis. It is worth noting that for a given dimension the measurement positions are fixed. Thus, the camera like detector can be easily changed to an array of Avalanche Photo Diodes (APD's).

Previously to estimate the probabilities of projection from the interference pattern described by \eqref{eq:interference-pattern-minima}, an initial calibration must be performed in order to compensate the diffraction envelope originated by the finite size of the slits. To this end, a mask of $d$ slits with equal intensities is displayed in both SLMs. On the other hand, as it was previously described, the canonical basis can not be measured by employing the same multiplexing method because this particular basis does not have an equal amplitude in each slit. Alternatively the $A(x)$ distribution is imaged onto SLM2 and simultaneously a phase grating, with different  period for each slit, is programmed on this modulator. The grating orientation is selected in such a way that the projection information on the canonical basis is multiplexed along the transverse direction of the interference pattern. This procedure is described in detail in the Supplementary Documents.
In his way, the proposed method allows us to reconstruct any state from only $d+1$ image acquisitions.

\begin{figure}[!ht]
\centering
\includegraphics[width=.42\textwidth]{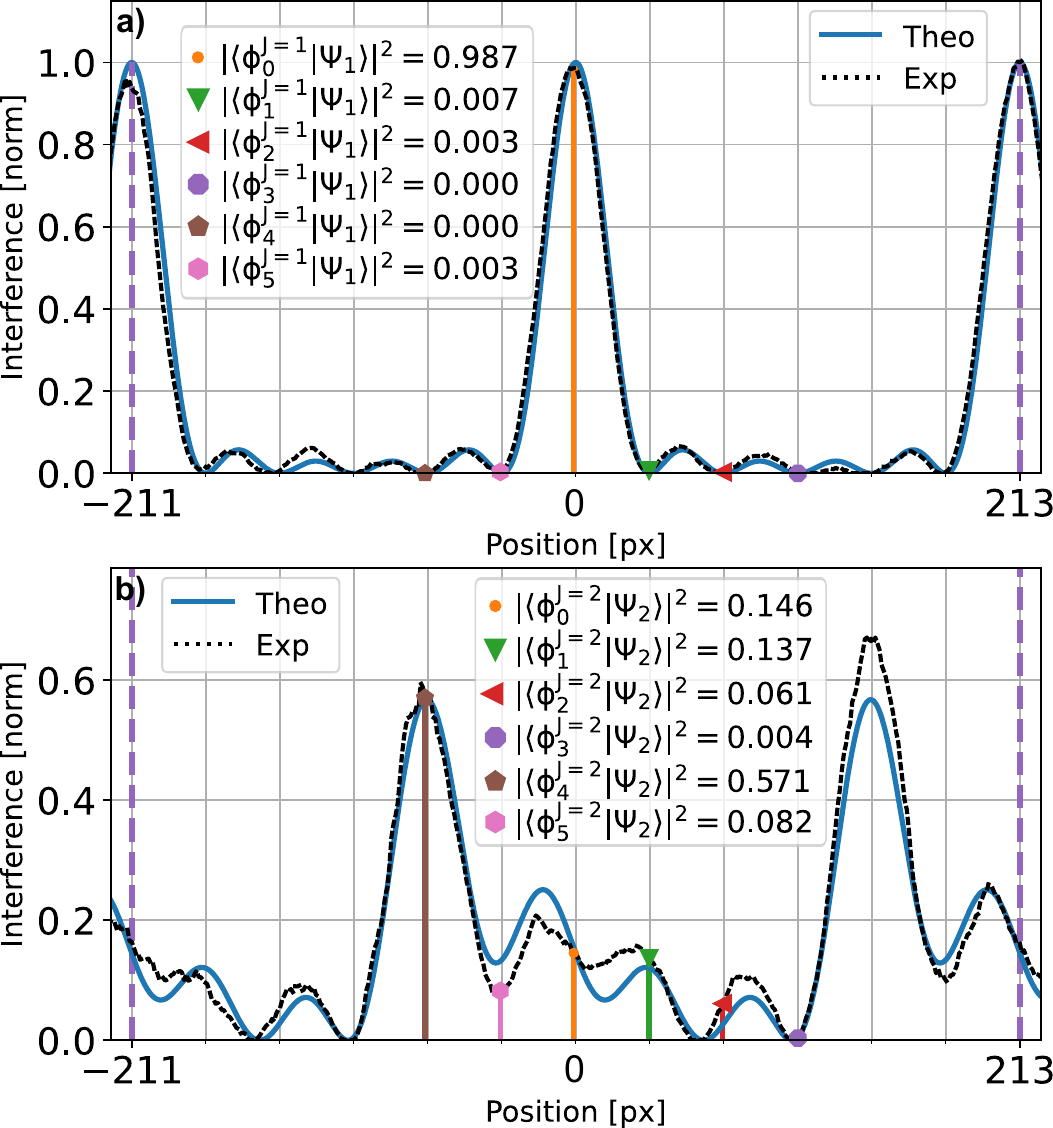}
\caption{Comparison between the interferograms resulting from a numerical simulation of the whole optical process (solid lines) and those obtained experimentally (dashed lines), for two different states $|\Psi_{1,2}\rangle$, and two different bases of the informational complete set $\{\mathcal{B}_J\}_{J=0}^5$. 
The interference patterns shown in panel (a) and panel (b) correspond to the state $|\Psi_1\rangle$ projected onto the $J=1-$basis, and the state $|\Psi_2\rangle$ projected onto the $J=2-$basis, respectively. The legends show the estimated probabilities of projection on each element of the basis.\label{fig:interference-experimental-comparison}}
\end{figure}

\textit{Results and discussion.} We prepared and reconstructed over 15 states of dimension $d=6$ for which the reconstruction fidelities are all over $0.970$. The complete set of $d+1$ bases used to perform these tomographies can be seen in the Supplementary Documents. Furthermore, for some of these states we performed the traditional projection scheme \cite{Lima2011}, where each of the $d \cdot (d+1) = 42$ elements of the $(d + 1)$ bases is programmed in SLM2, and the intensity is measured only at the center of the interference pattern. Under this measurement technique, the values of the fidelities are similar to those resulting from the multiplexed method. In Figure \ref{fig:interference-experimental-comparison} we show examples of the resulting interference patterns for two different states and projecting bases, comparing the theoretical result (solid lines) and the experimental result (dashed lines). Figure \ref{fig:interference-experimental-comparison}a shows the interference of the state \stateZeroDefAlt\ when projected on the $J=1-$basis. There is an excellent agreement between the expected and measured probabilities, and when fully reconstructed this state has a fidelity $F_1 = 0.980$. Figure \ref{fig:interference-experimental-comparison}b shows the interference of the state \stateThreeDef   when projected on the $J=2-$basis. This case also displays a good agreement between the expected and measured probabilities, and when fully reconstructed this state has a fidelity $F_2 = 0.974$.

\begin{figure}[!ht]
\centering
\includegraphics[width=.425\textwidth]{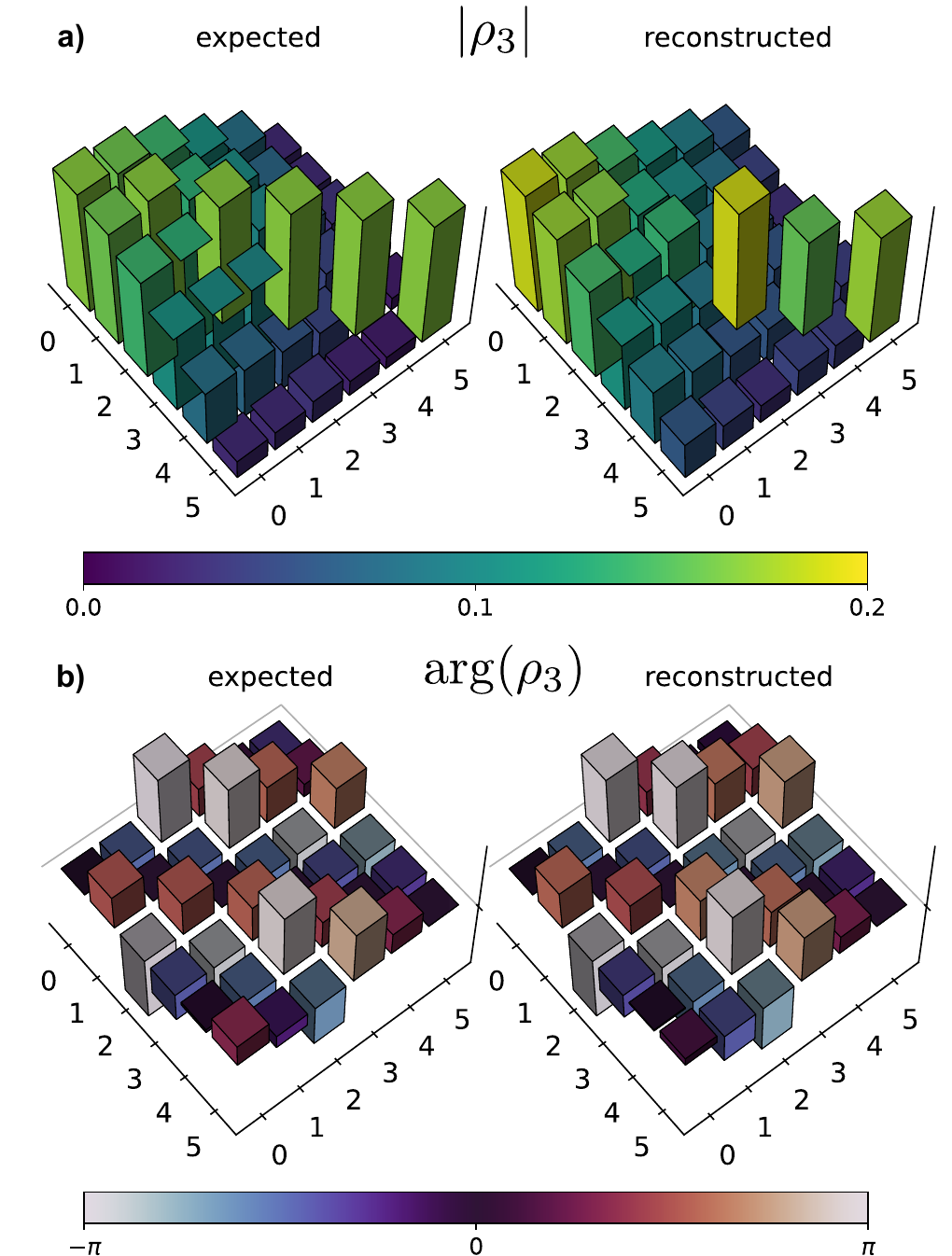}
\caption{Comparison between the theoretical (left column) and experimental (right column) of the density matrix of the state $\rho_3$, with a reconstruction fidelity of $F_3 = 0.984$.  a) Absolute value. b) Complex argument \label{matrices}}
\end{figure}

This tomographic method is completely general and can be used to reconstruct not only pure states but also mixed states. Following the recipe shown in Ref.~\cite{Varga2017}, we simulate a mixed state by combining, over time, pure states with identical real amplitudes and randomly varying relative phases. Specifically, we displayed on the SLM1 a sequence of masks representing the state
\begin{eqnarray}
|\Psi_3\rangle = \frac{1}{\sqrt{d}} \sum_{\ell} e^{i 2 \pi \ell/d} e^{i \Delta_\ell} |\ell\rangle,  
\end{eqnarray}
 where the $\Delta_\ell$ values are random variables uniformly distributed between 0 and $2 \pi\ell /d$. The projection probability on each of the bases $\{\mathcal{B}_J\}_{J=0}^d$ is measured following the method proposed in this article, that is, the intensity of the interference pattern in the selected positions is registered for each of the 7 masks prepared in SLM2. These intensities are averaged over the sampled random states $|\Psi_3\rangle$. 

The mixed state that results of randomly selecting $|\Psi_3\rangle$ is 
\begin{eqnarray}
\nonumber
\rho_3 & = &  \frac{1}{d} \sum_{\ell\neq\ell'} e^{i 3 \pi \ell/d} \mathrm{sinc}(\ell/d)\mathrm{sinc}(\ell'/d) |\ell\rangle\langle \ell'| +\\
& + & \frac{\delta_{\ell,\ell'}}{d} |\ell\rangle\langle \ell|.
\end{eqnarray}

Figure \ref{matrices} shows the comparison between the expected density matrix for $\rho_3$ (left column) and the experimentally reconstructed matrices (right column), with the upper row (a) showcasing the absolute value of these matrices and the lower row (b) showcasing the complex argument of these matrices. The agreement is excellent, with a reconstruction fidelity of $F_3 = 0.984$. Thus, even for a general mixed state, only $d+1$ measurement configurations are needed to reconstruct the state, while without multiplexing \cite{Lima2011, Varga2017} at least $d\cdot(d+1)$ configurations are needed (in the present case, this corresponds to 7 vs 42 measurement configurations).

In summary, we have presented a tomographic method that, by parallelizing projective measurements, allows for the reconstruction of an arbitrary $d-$dimensional photonic qudit state encoded in the discretized transverse momentum of photons. 
Since certain schemes enable the coupling of path and orbital angular momentum (OAM)~\cite{fickler2014}, the scope of potential applications of the method presented in this work, could be extended to OAM states.
Furthermore, we showed how to find an informationally complete set of bases that enable a fully multiplexed projective measurement implementation, highlighting the relationship between projective measurements and interference patterns. Thus, only $d+1$ experimental settings are required to obtain the density matrix of any unknown qudit state, implying a reduction from $O(d^2)$ to $O(d)$.

       
\section*{Acknowledgments}
This work was supported by Universidad de Buenos Aires [UBACyT 20020170100564BA] and ANPCyT [PICT-2020-SERIEA-02031]. QPS acknowledges support from IKUR Strategy under the collaboration agreement between Ikerbasque Foundation and DIPC/MPC on behalf of the Department of Education of the Basque Government.

\bibliography{bibliography}

\onecolumngrid
\section{Supplementary Document}
\subsection{Bases used for the $d=6$ tomography}
The first basis $\mathcal{B}_0$ is the canonical one. The second basis $\mathcal{B}_1$ is generated by the vector $|\Phi_0^{J=1}\rangle = \frac{1}{\sqrt{6}}\sum_{\ell=0}^{5}|\ell\rangle$. The remaining bases are generated by applying the method described in the main manuscript to randomly chosen first vectors. The random selection is performed several times, and the set of bases with the lower condition number is chosen. A list with the first vectors used is shown below (real and imaginary part of the complex coefficients rounded to the third decimal place):
\begin{eqnarray}
|\Phi_0^{J=1}\rangle \approx & 
(.408 + .000i)|0\rangle + (.408 + .000i)|1\rangle + (.408 + .000i)|2\rangle + \\  \nonumber
+& (.408 + .000i)|3\rangle +  (.408 + .000i) |4\rangle +  (.408 + .000i) |5\rangle\\
|\Phi_0^{J=2}\rangle  \approx & (.203 + .354i)|0\rangle + (.043 + .406i)|1\rangle + (.010 + .408i)|2\rangle + \\\nonumber
+& (.396 - .100i)|3\rangle +  (0.406 + 0.042i) |4\rangle +  (-0.302 + 0.274i) |5\rangle\\
|\Phi_0^{J=3}\rangle \approx &
(-.400 + .082i)|0\rangle + (-.359 + .194i)|1\rangle + (.301 - .276i)|2\rangle + \\\nonumber
+& (-.408 + .016i)|3\rangle +  (.380 + .148i) |4\rangle +  (.364 - .185i) |5\rangle\\
|\Phi_0^{J=4}\rangle \approx &
(-.218 - .345i)|0\rangle + (-.001 - .408i)|1\rangle + (-.406 + .043i)|2\rangle + \\\nonumber
+& (-.015 + .408i)|3\rangle +  (  .038 + .407i) |4\rangle +  (  -.332- .238i) |5\rangle\\
|\Phi_0^{J=5}\rangle \approx &
(-.392 + .115i)|0\rangle + (.071 - .402i)|1\rangle + (.034 + .407i)|2\rangle + \\\nonumber
+& (.209 - .351i)|3\rangle +  (.121 - .390i) |4\rangle +  (-.402 + .070i) |5\rangle\\
 |\Phi_0^{J=6}\rangle \approx &
(.149 + .380i)|0\rangle + (-.037 + .407i)|1\rangle + (.076 - .401i)|2\rangle + \\\nonumber
+& (.352 - .207i)|3\rangle +  (.408 + .012i) |4\rangle +  (-.050 - .405i) |5\rangle
\end{eqnarray}

\section{Mask for multiplexed measurement on the canonical basis}
\begin{figure}[h]
\centering
\fbox{\includegraphics[width=\linewidth]{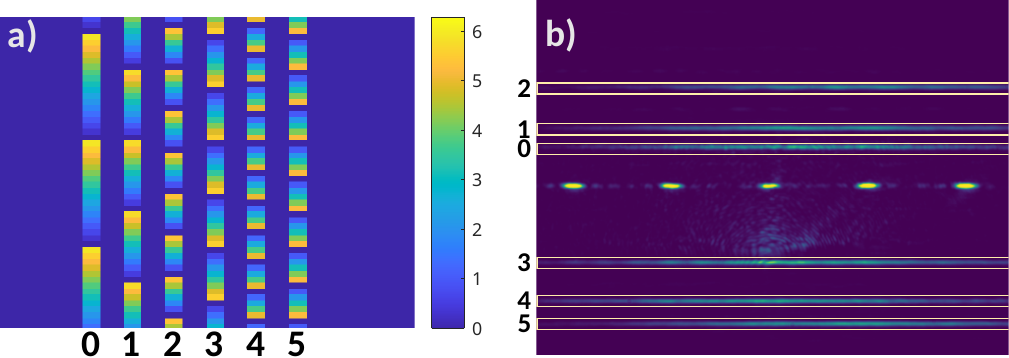}}
\caption{a) Shows the phase diffraction gratings programmed on SLM2 to perform spatial multiplexing of the projections on the canonical basis. b) Shows the position measurements for each slit, and the intensity in normalized units. The slit interference on the 0th order is due to the inefficiencies of the mask.}
\label{fig:false-color}
\end{figure}

As it was mentioned in the main article, the canonical basis does not fulfil the condition $|b_\ell|^2 =1/d$ for $\ell=0,\cdots,d-1$, and thus it can not be multiplexed in positions along the interference pattern. However, an alternate strategy can be devised to multiplex this basis for this case. The $A(x)$ distribution that represents the state to be reconstructed is imaged onto SLM2 and simultaneously a phase grating, with different  period for each slit, is programmed on this modulator. Figure S1a shows the phase map programmed in SLM2 that performs this encoding. Each of the slits features a blazed diffraction grating with different spatial periods. Figure S1b shows the resulting far field and the multiplexed positions for each of the element of the basis. To account for the different efficiencies of the slits, resulting for the pixel discretization of the the blaze grating, an initial calibration measurement is performed in this basis.

\end{document}